\documentclass[numbers]{aipproc}
\layoutstyle{8x11double}

\usepackage{amsmath}    

\newcommand{\beq}{\begin{equation}}
\newcommand{\eeq}{\end{equation}}
\newcommand{\bqa}{\begin{eqnarray}}
\newcommand{\eqa}{\end{eqnarray}}

\def\square{\vcenter{\vbox{\hrule height.4pt
          \hbox{\vrule width.4pt height4pt
          \kern4pt\vrule width.3pt}\hrule height.4pt}}}

\title{Kaon condensation in CFL quark matter, the Goldstone theorem, and the 2PI Hartree approximation}
\author{Lars E. Leganger}{address={Department of Physics,
       Norwegian University of Science and Technology, H\o gskoleringen 5,
       N-7491 Trondheim,
       Norway}}

   \begin{abstract}
At very high densities, QCD is in the color-flavor-locked phase, which
is a color-superconducting phase.
The diquark condensates break chiral symmetry in the same way as it
is broken in vacuum QCD and
gives rise to an octet of pseudo-Goldstone bosons and a superfluid mode.
The lightest of these are the charged and neutral kaons. 
For energies below the superconducting gap, 
the kaons are described by an 
$O(2)\times O(2)$-symmetric effective scalar field theory with 
chemical potentials. 
We use this effective
theory to study 
Bose-condensation of kaons and their 
properties as functions of the temperature and the chemical potentials.
We use the 2-particle irreducible effective action formalism in the
Hartree approximation. The renormalization of
the gap equations and the effective potential is studied in detail
and we show that the counterterms are independent of temperature and
chemical potentials.
We determine the
phase diagram and the medium-dependent quasiparticle masses.
It is shown that the Goldstone theorem is satisfied to a very good
approximation.
        \end{abstract}

\begin{document}

\maketitle

\section{Introduction}

At high density and low temperature, we know 
that QCD is in the color-flavor-locked (CFL) 
phase~\cite{frankie,frank,raja,mark}. 
This state is a color
superconducting state since the quarks form Cooper pairs 
as electrons in an ordinary superconductor. 
The attraction between the quarks, which renders the Fermi surface 
unstable against the formation of Cooper pairs, is provided by one-gluon
exchange.

At asymptotically high densities, one can ignore the strange-quark mass
and quarks of all three colors and all three flavors participate in
a symmetric manner in the pairing.
The original symmetry group 
$SU(3)_{\rm c}\times SU(3)_L\times SU(3)_R \times U(1)_B$ 
is broken down
to $SU(3)_{c+L+R}$ which is a linear combination of the generators of the
original group. This locks rotation in color space with
rotations in flavor space and this has given the name to the phase.
Since the symmetry-breaking pattern is the same as in vacuum QCD, 
the low-energy properties 
of the CFL phase can be described in terms of an effective chiral Lagrangian
for the octet of (pseudo-) Goldstone modes and the superfluid
mode~\cite{gatto,sonny,kaon,kaon2,sjafer}.
An important difference between chiral perturbation theory
in the vacuum and in the CFL phase is that the latter is at high density
and the Lagrangian is therefore coupled to chemical potentials via
the zeroth component of a "gauge field". 

At asymptotically
high densities, all modes are exactly massless since one can
ignore the quark masses. 
At moderate densities, this is no longer the case.
The quark masses cannot be neglected and chiral symmetry is explicitly
broken. This implies that only the superfluid mode is exactly
massless, while 
the other mesonic modes acquire masses. 
This is relevant for the interior of a neutron star.
In this case, the quark
chemical potential is of the order of 500 MeV, while the strange quark mass
is somewhere between the current quark mass of approximately 100 MeV
and the constituent quark mass of approximately 500 MeV~\cite{alfordus}.
The mass spectrum in the CFL is the opposite of vacuum QCD and
the lightest massive modes are expected to be the charged and neutral kaons
$K^+/K^-$ and $K^0/\bar{K}^0$.

The 2PI effective action formalism~\cite{cjt} has recently
been used to study the thermodynamics
of pions and  kaons and their condensation.
In Ref.~\cite{alfordus}, the authors 
applied the 2PI effective action formalism 
in the Hartree approximation to an effective $O(2)\times O(2)$-symmetric
scalar field and calculated the phase diagram
and the critical temperature for Bose-condensation of kaons.
The effects of imposing
electric charge neutrality were also investigated.
The scalar theory for the kaons were derived
from the effective chiral 
Lagrangian, where the parameters depend on the baryon chemical potential.
Renormalization issues were not addressed.

In the present paper we reconsider the problem of kaon condensation
from a somewhat different angle, and consider the renormalization of the theory. In particular, we find that all
divergences in the gap equations and the effective potential can be 
eliminated by counterterms that are independent of temperature and chemical
potentials. We also show that the violation
of Goldstone's theorem is negligible. A more detailed presentation of the calculations can be found in \cite{Andersen:2008tn}.

\section{Kaons in the CFL phase}

The chiral effective Lagrangian of dense QCD in the CFL phase is given by~\cite{sonny}
\bqa
{\cal L}&=&{1\over4}f_{\pi}^2{\rm Tr}
\left[
\left(\partial_0\Sigma+i[A,\Sigma]\right)
\left(\partial_0\Sigma-i[A,\Sigma]^{\dagger}\right)
\right.\nonumber \\ &&\nonumber\left.
-v_{\pi}^2(\partial_i\Sigma)(\partial_i\Sigma^{\dagger})
\right] 
\\ &&
+{1\over2}af_{\pi}^2\det M{\rm Tr}[M^{-1}(\Sigma+\Sigma^{\dagger})]\;+\cdots,
\eqa
where $f_{\pi}$, $v_{\pi}$, and $a$ are constants,
the meson field $\Sigma=e^{i\lambda^a\phi^a/f_{\pi}}$,
where $\lambda^a$ are the Gell-Mann matrices and $\phi^a$ describe
the octet of Goldstone bosons.
The matrix $A=\mu_QQ-{M^2\over2\mu}$ acts as the zeroth component of a gauge field,
where $\mu_Q$ is the chemical potential for electric charge $Q$, $\mu$
is the baryon chemical potential, $Q={\rm diag}(2/3,-1/3,-1/3)$, and
$M={\rm diag}(m_u,m_d,m_s)$.
At asymptotically high densities, one can use perturbative QCD calculations
to determine the parameters 
by 
matching~\cite{sonny,kaon2,sjafer}. For moderate densities, which are
relevant for compact stars, a precise determination of the parameters is difficult.

Expanding to fourth order in the meson fields, 
one
obtains an effective Lagrangian for the kaons
where the parameters depend on the quark masses, chemical potentials, and $f_\pi$, as determined by matching \cite{alfordus}. Generally these parameters must satisfy some RG
equations, but we do not know how they evolve down to values of the chemical
potential relevant for a star. We therefore chose to use our effective theory in a more
traditional way, where we are content with using coupling constants {\it inspired} by the matching.

The kaons are 
written as a complex doublet, 
$(K^0,K^+)=(\Phi_1,\Phi_2)$. 
The Euclidean Lagrangian with an $O(2)\times O(2)$ symmetry is given by
\bqa\nonumber
{\cal L}&=&
\left[(\partial_{0}+\mu_0)\Phi_1^{\dagger}\right]\left[(\partial_{0}-\mu_0)\Phi_1\right]
+(\partial_{i}\Phi_1^{\dagger})(\partial_{i}\Phi_1) \\ \nonumber & & 
+ \left[(\partial_{0}+\mu_+)\Phi_2^{\dagger}\right]\left[(\partial_{0}-\mu_+)\Phi_2\right]
\\ \nonumber &&
+(\partial_{i}\Phi_2^{\dagger})(\partial_{i}\Phi_2) 
+m_0^2\Phi^{\dagger}_1\Phi_1+m_+^2\Phi^{\dagger}_2\Phi_2
\\&& \nonumber
+{\lambda_0\over2}\left(\Phi^{\dagger}_1\Phi_1\right)^2
+{\lambda_+\over2}\left(\Phi^{\dagger}_2\Phi_2\right)^2 
\\&&
+{\lambda_H}\left(\Phi^{\dagger}_1\Phi_1\right)\left(\Phi^{\dagger}_2\Phi_2\right)
\;.
\label{lag2x2}
\eqa
The chemical potentials
$\mu_0$ and $\mu_+$ are associated with the two conserved charges for
each complex field $\Phi_i$, related to the 
quark chemical potentials by
$
\mu_0=\mu_d-\mu_s \;,
\mu_+=\mu_u-\mu_s \;. 
$

In order to allow for a condensate of neutral kaons, we introduce
an expectation value $\phi_0$ for $\Phi_1$ and write
\bqa
\Phi_1&=&
{1\over\sqrt{2}}\left(\phi_0+\phi_1+i\phi_2\right)
\;,
\eqa
where $\phi_1$ and $\phi_2$ are quantum fluctuating fields.
The inverse tree-level propagator can be written as a block-diagonal 
$4\times4$ matrix $D_0^{-1}={\rm diag}\left(D_a,D_b\right)$:
\bqa
D_a=
\begin{small}
\left(\begin{array}{cc}
P_n^2+m_1^2 - \mu_0^2&-2\mu_0\omega_n
\\
2\mu_0\omega_n&P_n^2+m_2^2 - \mu_0^2
\end{array}\right)
\end{small}
\;,\\
D_b=
\begin{small}
\left(\begin{array}{cc}
P_n^2+m_3^2 - \mu_+^2&-2\mu_+\omega_n
\\
2\mu_+\omega_n&P_n^2+m_3^2 - \mu_+^2 \\
\end{array}\right)
\end{small}\;,
\eqa
where $P_n^2=\omega_n^2+p^2$ and the tree-level masses are
$
m_1^2=m_0^2+{3\lambda_0\over 2}\phi_0^2,
m_2^2=m_0^2+{\lambda_0\over 2}\phi_0^2,
$ and $
m_3^2=m_+^2+{\lambda_H\over 2}\phi_0^2\;.
$

The 2PI effective action is given by
\bqa
\Omega[\phi_0,D]&=&
{1\over2}
\left(m_0^2-\mu_0^2\right)\phi_0^2
+{\lambda_0\over8}\phi_0^4 
+{1\over2}{\rm Tr}\ln D^{-1}
\nonumber\\&&
+{1\over2}{\rm Tr}D_0^{-1}D
+\Phi[D]\;,
\eqa
where $\Phi[D]$ contains all 2PI vacuum diagrams.
In the Hartree approximation, we include all double-bubble diagrams
which can be written in terms of $O(2)\times O(2)$ invariants~\cite{fejos}
\bqa
\Phi[D]&=&{\lambda_0\over8}
\left[{\rm Tr}\left(D_a\right)]^2
+2{\rm Tr}\left(D_a^2\right)\right]
+
{\lambda_+\over8}
\left[{\rm Tr}\left(D_b\right)]^2
\right.\nonumber\\&&\left.
+2{\rm Tr}\left(D_b^2\right)\right]
+{\lambda_H\over4}({\rm Tr}D_a)({\rm Tr}D_b)\;.
\eqa
The contributions from the sum-integrals can be split into divergent medium-independent
and convergent medium-dependent parts, allowing medium-independent counterterms.
After renormalization, the self-consistent set gap equations
read
\bqa
\label{m12pi}
\Delta M^2_1 &=&
{1\over2}\left[
3\lambda_0{J}^{c,T}_1 + \lambda_0{J}^{c,T}_2 + 2\lambda_H{J}^{c,T}_3\right]
\;,
\\
\label{m22pi}
\Delta M^2_2 &=& {1\over 2}\left[\lambda_0{J}^{c,T}_1 + 3\lambda_0{J}^{c,T}_2 + 2\lambda_H{J}^{c,T}_3\right]
\;,
\\
\Delta M^2_3 &=& {1\over2}\left[
\lambda_H{J}^{c,T}_1 + \lambda_H{J}^{c,T}_2 + 4\lambda_+{J}^{c,T}_3\right]\;,
\\
0 &=& \phi_0\left[m_0^2-\mu_0^2+{\lambda_0\over
2}\phi_0^2+{1\over2}\left[3\lambda_0{J}^{c,T}_1
\right.\right.\nonumber\\&&\left.\left.+ \lambda_0{J}^{c,T}_2 +
2\lambda_H{J}^{c,T}_3\right]\right]\;,
\label{phi2pi}
\eqa
where $\Delta M^2_i = M^2_i-m_i^2$ and $J$'s are medium-dependent convergent integrals involving the dressed masses $M$. 

The renormalized effective potential is given by
\bqa\nonumber
\Omega&=&{1\over2}(m^2_0-\mu_0^2)\phi_0^2+{\lambda_0\over8}\phi_0^4
+ {1\over2}{\cal J}^{c,T}_1
+ {1\over2}{\cal J}^{c,T}_2
\\ \nonumber & & 
+ {\cal J}^{c,T}_3
-\frac{1}{2}\Delta M_1^2{J}^{c,T}_1
-\frac{1}{2}\Delta M_2^2{J}^{c,T}_2
-\Delta M_3^2{J}^{c,T}_3\\ \nonumber
&&
+ {3\lambda_0\over 8}
\left({J}^{c,T}_1\right)^2
+ {3\lambda_0\over 8}\left({J}^{c,T}_2\right)^2
+\lambda_+\left({J}^{c,T}_3\right)^2 
\\ & &
+ {\lambda_0\over 4}
{J}^{c,T}_1{J}^{c,T}_2 
+{\lambda_H\over 2}
{J}^{c,T}_1 {J}^{c,T}_2 
+ 
{\lambda_H\over 2}
{J}^{c,T}_2{J}^{c,T}_3 \;.
\nonumber\\&&
\label{finalomega}
\eqa

\section{Results}

\begin{figure}[htb]
  \resizebox{1\columnwidth}{!}{\includegraphics{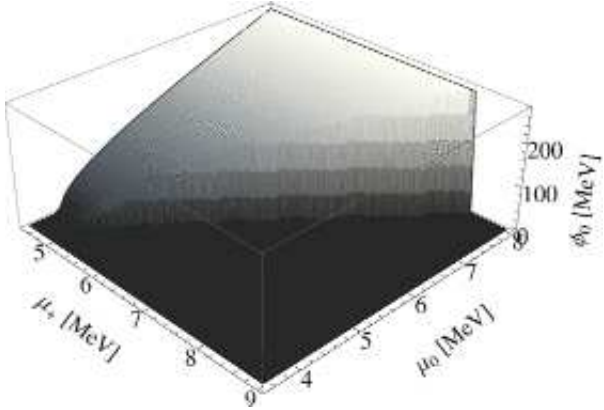}}
\caption{Neutral
kaon condensate as a function of the chemical potentials
$\mu_0$ and $\mu_+$ for $T=0$.}
\label{cond1}
\end{figure}

In Fig.~\ref{cond1}, we show
the neutral kaon condensate as a function of $\mu_0$ and $\mu_+$ for $T=0$.
For $\mu_+=0$, i.e. along the $\mu_0$-axis, 
there is a second-order phase transition to a neutral phase with a
kaon condensate at a critical chemical potential $\mu_0=m_2$. 
This is the CFL-$K^0$ phase.
For larger values of $\mu_+$ the transition becomes first order, to a phase with 
condensate of charged kaons. This CFL-$K^+$ phase condensate is not
shown in the figure.
Thus there is a competition between the neutral
and the charged condensates and nowhere do they exist simultaneously.
The transitions to the kaon-condensed phases are density driven.
In \cite{Andersen:2008tn} we also examine effects of imposing electric charge neutrality.

\begin{figure}[htb]
  \resizebox{0.9\columnwidth}{!}{\includegraphics{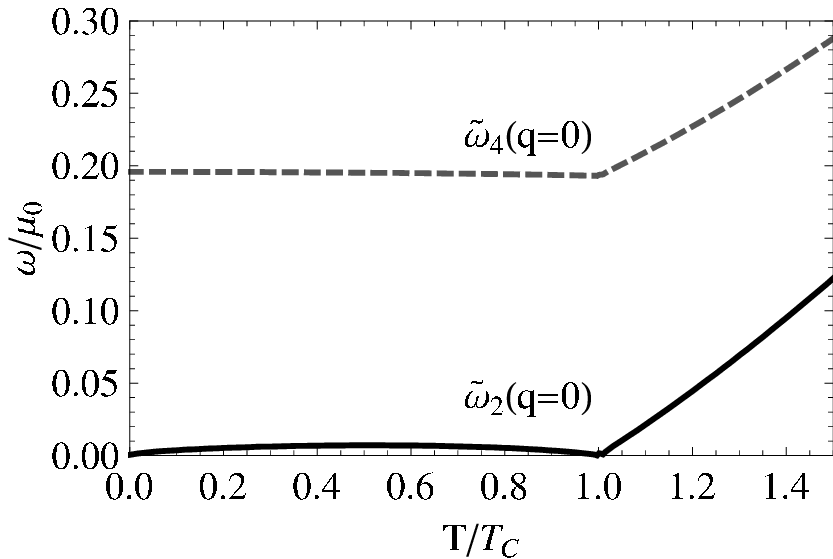}}
\caption{Mass gaps for the $K^+$ and $K^0$ modes
for $\mu_0=\mu_+=4.5$ MeV and as a
function of $T$ normalized to $T_c$.}
\label{energygaps}
\end{figure}

In Fig.~\ref{energygaps} we show the masses of $K^0$ ($\omega_2(q=0)$) 
and $K^+$ ($\omega_4(q=0)$) for $\mu_0=\mu_+=4.5$  
MeV and as functions of $T$ normalized to $T_c \sim 120$ MeV. We notice that the
mass of $K^0$ is not strictly zero, which  
explicitly shows that the Goldstone theorem is
not respected by the Hartree approximation. In Ref.~\cite{alfordus}, the authors make
some further approximations of the sum-integrals appearing in the
gap equations. These approximations give rise to an exactly gapless mode.
As pointed out in their paper and as can be seen in 
Fig.~\ref{energygaps}, this is a very good approximation.

\section{Summary and outlook}

We have studied the phase
diagram and the quasiparticle masses of the $O(2)\times O(2)$ model 
where the neutral kaons condense at sufficiently low temperature and
sufficiently large value of the chemical potential, and showed it is possible to renormalize the gap equations and effective potential in a medium-independent way.
If the transition is first order, it turns out that the transition is not
to a symmetric state but to a state with a $K^+$ condensate.
This is in agreement with the findings of Alford, Braby,
and Schmitt~\cite{alfordus}.

One drawback of the 2PI Hartree approximation is that it does not
obey Goldstone's theorem. 
For practical purposes we find the 
violation is negligible which is reassuring. 
We therefore believe that the 2PI Hartree approximation is a useful
nonperturbative approximation for systems in thermal equilibrium.

The Hartree approximation and the large-$N$ limit are both mean-field
approximations. It would be desirable to go beyond mean field
for example by including next-to-leading corrections in the $1/N$-expansion.

\end{document}